\documentclass[journal]{IEEEtran}

\usepackage{graphicx}
\ifCLASSINFOpdf

\else

\fi
\usepackage{caption}
\usepackage{subcaption}
\usepackage{amsmath}
\usepackage{amssymb}
\usepackage{tabularx}
\usepackage{color}
\usepackage{soul}
\usepackage{comment}
\usepackage[font=small,labelfont=bf]{caption}
\usepackage{hyperref}
\usepackage{array}
\usepackage{lipsum}
\usepackage[font={footnotesize}]{caption}
\usepackage{bm}
\usepackage{amsmath}
\usepackage{upgreek}
\usepackage{physics}
\usepackage{esvect}
\usepackage{color, soul}
\usepackage{cite}
\newcolumntype{?}{!{\vrule width 1pt}}
\usepackage{amssymb}
\usepackage{dblfloatfix}

\begin{document}

\title{Ingress Cryogenic Receivers Toward Scalable Quantum Information Processing: Theory and System Analysis}

\author{Malek~Succar,~\IEEEmembership{Graduate Student Member,~IEEE,}
        Mohamed I.~Ibrahim,~\IEEEmembership{Member,~IEEE}
\thanks{M. Succar and M. Ibrahim are with the Department
of Electrical and Computer Engineering, Cornell University, Ithaca,
NY, 14853, USA (ms3622@cornell.edu and mibrahim@cornell.edu).}
}
\maketitle

\begin{abstract}
Current control techniques for cryogenically cooled qubits are realized with coaxial cables, posing multiple challenges in terms of cost, thermal load, size, and long-term scalability. Emerging approaches to tackle this issue include cryogenic CMOS electronics at 4~K, and photonic links for direct qubit control. In this paper, we propose a multiplexed all-passive cryogenic high frequency direct detection control platform (cryo-HFDD). The proposed classical interface for direct qubit control utilizes optical or sub-THz bands. We present the possible tradeoffs of this platform, and compare it with current state-of-the-art cryogenic CMOS and conventional coaxial approaches. We assess the feasibility of adopting these efficient links for a wide range of microwave qubit power levels. Specifically, we estimate the heat load to achieve the required signal-to-noise ratio $SNR$ considering different noise sources, component losses, as well as link density. We show that multiplexed photonic receivers at 4~K can aggressively scale the control of thousands of qubits. This opens the door for low cost scalable quantum computing systems.

\end{abstract}

\begin{IEEEkeywords}
Qubit, Heat Load, Photonic Link, Sub-THz Link, All-Passive
\end{IEEEkeywords}

\IEEEpeerreviewmaketitle

\section{Introduction}
\IEEEPARstart 
{A}{ddressing} thousands to millions of superconducting or spin qubits is proving to be a challenging task due to the complexity of maintaining coaxial lines within the cooling power and size of a typical dilution fridge.  
Conventional methods to control and readout cryogenically cooled qubits use coaxial cables, which limits system scaling beyond a few hundreds of qubits, motivating researchers to implement novel control interfaces. 
Complementary metal-oxide-semiconductor (CMOS) integrated circuits stand out as a promising solution because of their scalability and reduced cost, size, and power. 
Cryogenic CMOS (cryoCMOS) circuitry operating at 4~K has been demonstrated recently to interface with a few superconducting or spin qubits working at milliKelvin (mK) temperatures~\cite{Google, IBM, jssc_drag, Intel, review_paper}. 
However, these cryoCMOS pulse modulation controllers still suffer from relatively high power consumption (a few mW/qubit at 4~K).

\begin{figure}[!htp]
    \includegraphics [scale=0.65]{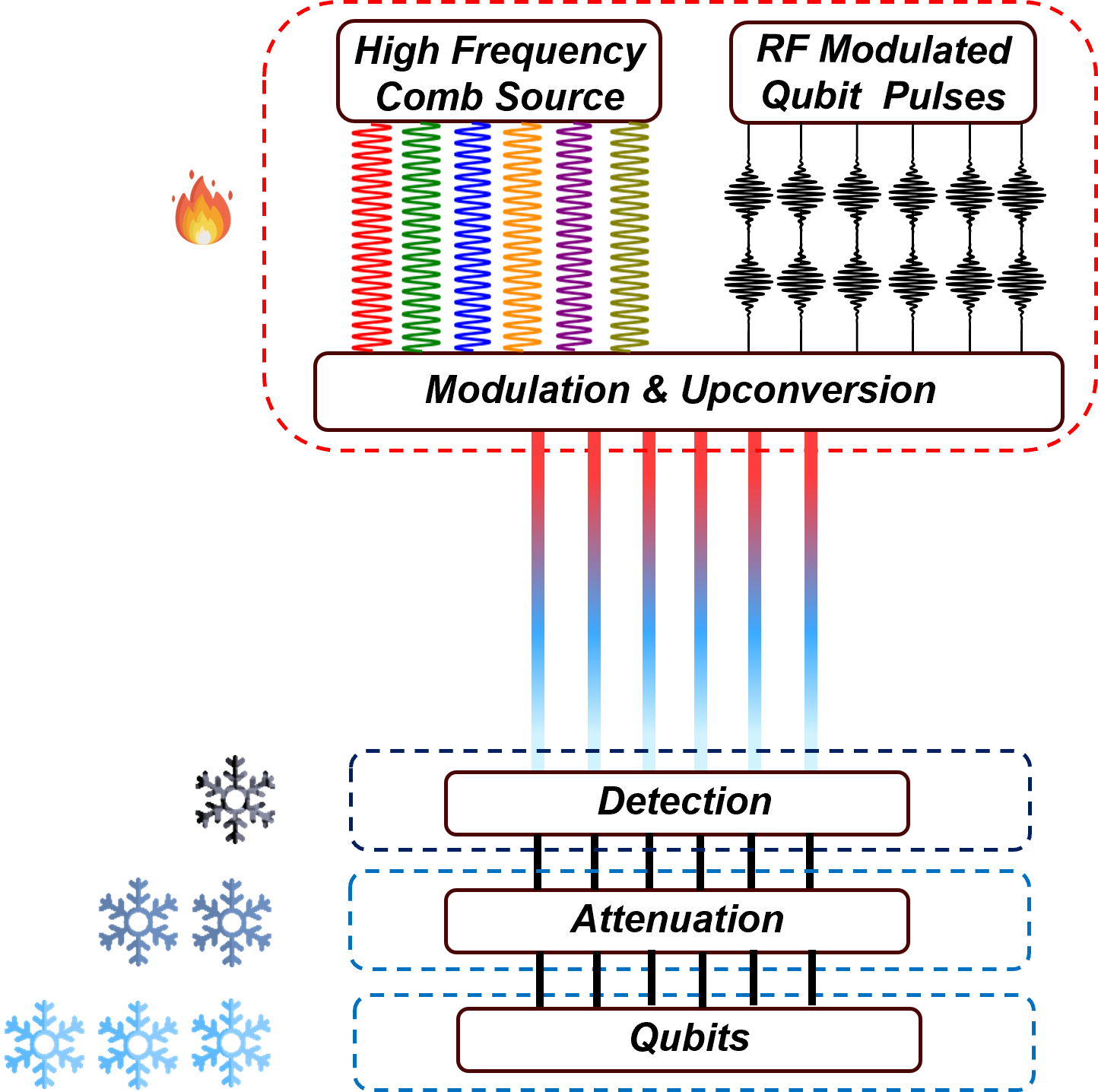}
    \centering
    \captionof{figure}{Cryogenic High Frequency Direct Detection (cryo-HFDD) qubit control platform.}
    \label{concept}
\end{figure} 

Other techniques have been proposed to overcome that limit. 
Photonic receivers for direct qubit control emerge as a potential solution to achieving scalability~\cite{NIST, NIST_2, Amir, warner2025coherent, Washington}.
For example, in ~\cite{NIST}, qubit pulses are sent to cryogenic temperatures using a photonic link. 
Off-the-shelf bulk modified uni-traveling carrier (MUTC) InGaAs photodiodes with an extended bandgap at 20 mK are used to downconvert the qubit pulses.
However, a detailed study is yet to be done to assess the trade-offs of this approach (heat load, attenuation, size, etc.), which is essential in determining a scalable path for millions of qubits. 
On the other hand, bi-directional terahertz (THz) links at 4~K that wirelessly deliver classical digital control/readout signals of quantum systems to small cryogenic fridges were also demonstrated in ~\cite{THzBackscatter, wang2025wireless}.
4~Gbps uplink and downlink operation with 176~fJ/b and 34~fJ/b, respectively, were achieved. 
However, this approach has not yet been explored to scale qubit control by directly sending RF pulses. Sub-THz dielectric waveguides made from nonconductive non-polar polymers, such as the commercially available Polytetrafluorethylene (PTFE or Teflon) are promising alternatives.
These waveguides are characterized by their low dielectric loss tangent at the mm-wave range \cite{afsar2012precision}.

In this manuscript,  we introduce a multiplexed all-passive cryogenic high frequency direct detection (cryo-HFDD) scheme, as shown in Fig.~\ref{concept}. 
The proposed ingress interfaces for direct qubit control can take advantage of high frequency bands such as sub-THz (140~GHz-220~GHz) and optical communication ($\sim$1550~nm) bands to carry modulated RF pulses over dielectric waveguides (fibers) inside a dilution fridge to the qubits, and thus reduce passive heat load and size of qubit control lines.
The signal then gets downconverted in the fridge using peak detectors, followed by passive matching networks to act as a filter, thereby extracting the qubit RF control pulse and optimizing the detector load for optimal signal-to-noise ratio ($SNR$). 
We present a detailed system analysis to determine the required heat load to meet the $SNR$, taking into account factors such as receiver (RX) operating temperature, system attenuation, component losses, detector responsivity ($\mathcal{R}$), and cable size. 
In Section II, we discuss the link passive thermal load and the $SNR$ specifications for typical superconducting qubits. 
We then introduce, in Section III, the limitations of photonically driving qubits at the mixing chamber (MXC) stage and discuss the possibility of control from the 4~K stage. 
In Section IV, we compare this approach to sub-THz links. In Section V, we discuss future scalability projections, and then Section VI concludes the manuscript.

\section{Link Passive Heat Load and $SNR$ Specification}
\label{section_passive_heat}

Conventional control of cryogenically cooled qubits is done using stainless steel coaxial cables from room temperature to the 4~K stage, followed by superconducting cables to the MXC stage~\cite{Attenuation}. These cables have a passive thermal load which can be calculated using Fourier's law for heat flow defined as:
\begin{equation}
        P_{i}={\displaystyle\int_{T_{i-1}}^{T_{i}}} dT{\displaystyle\frac{\rho_{0}(T)A_{o}+\rho_{d}(T)A_{d}+\rho_{c}(T)A_{c}}{L_i}}
        \label{fourier}
\end{equation}

\noindent $\rho_{o}$, $\rho_{d}$, $\rho_{c}$, amd $A_{o}$, $A_{d}$, $A_{c}$, are the thermal conductivities and cross sections of the outer conductor, dielectric, and center conductor of the coaxial cable, respectively, and $L_{i}$ is the cable length connecting temperature stages $i-1$ and $i$. Considering a typical 1m long stainless steel cable used in cryostats such as the UT-085-SS-SS, the heat dissipation at 4~K is $\sim$1.2~mW with 2~dB/m of attenuation \cite{Attenuation}. 
Assuming one coaxial cable from room temperature (RT) to 4~K is used for X/Y control of a single qubit, $\sim$ 1500 qubits at most can be controlled without exceeding the cooling power of the 4~K stage ($\sim$ 1.5~W). 
Therefore, using other communication channels with lower passive thermal load, size, and signal attenuation is crucial for highly dense qubit control. 
Specifically, dielectric waveguides (fibers) are attractive candidates to minimize passive heat load.
As shown in Table \ref{table_one}, an optical fiber made of 10~$\mu$m $SiO_2$ core diameter and 127 $\mu$m cladding diameter \cite{agrawal2012fiber}, has more than two orders of magnitude less heat load compared to the stainless steel cable, while being nearly lossless. 
A 1~mm$\times$1~mm Teflon dielectric waveguide in the range from 140~GHz-220~GHz, also benefits from reduced heat load. Teflon dielectric waveguides have an insertion loss of $\sim$5-10~dB/m at room temperature \cite{afsar2012precision}.
The 140 GHz-220 GHz range is optimal for maximized bandwidth and minimized loss of the dielectric waveguide.


\begin{table}[h!]
    \centering
\begin{tabularx}{0.5\textwidth} { 
  | >{\centering\arraybackslash}X 
  | >{\centering\arraybackslash}X 
  | >{\centering\arraybackslash}X | }
 \hline
\textbf{Specification} & \textbf{Fiber Optic Cable} & \textbf{Dielectric Waveguide} \\
 
 \hline
\textbf{Loss at 300~K}  & 0.2 dB/Km & $\sim$ 5 dB/m  \\
\hline
\textbf{Core Size}  & 10 $\mu$m diameter & 1~mm$\times$1~mm  \\
 \hline
\textbf{Heat Load at 50~K} & $\sim$ 6 $\mu$W  & $\sim$ 50 $\mu$W \\
\hline
\textbf{Heat Load at 4~K}  & $\sim$ 1 $\mu$W & $\sim$ 9 $\mu$W\\
\hline
\end{tabularx}\\ 
    \caption{Passive thermal load comparison for a 1m long cable}
    \label{table_one}
\end{table}

The total heat load is defined as $P_{total}=P_{passive} + P_{active}$ where $P_{passive}$ is the passive heat load dissipation of the control link, $P_{active}$ is the average active heat load consumed by the cryogenic receiver.
Since we focus on dielectric waveguide-based links as a scalable approach for direct qubit control, we ignore the passive heat load.
To determine the active heat load for the proposed cryo-HFDD interfaces, we define the required $SNR$ to achieve certain qubit power, which depends on the control pulse length and the $\ket{0}$ to $\ket{1}$ qubit transition frequency represented as $\omega_{01}$. 
The $SNR$ requirement depends on the power required to drive the qubit, and its operating temperature.   
For example, typical superconducting qubits operate at a few tens of mK, and require average power in the range of -60~dBm to -90~dBm~\cite{Google, IBM, Attenuation, Fluxonium}.
The receiver equivalent noise at the qubit stage should be smaller than $\hbar\omega_{01}$ \cite{9134701}, therefore we set this target to match the noise floor at the qubit operating temperature, where $\hbar$ is the reduced Planck constant. For example, for a qubit operating at 30~mK with average driving power of $P_{\mu W,30mK}$ of -80~dBm, the $SNR$ of the receiver needs to be 134 dB/Hz (i.e. noise floor at 30~mk is -214~dBm/Hz). 
Here, we consider three main cooling stages below the 50~K cooling plate, with the following temperatures: 4 K, 882 mK, and 30 mK \cite{Attenuation}, where in the following analysis we assume the qubit operating temperature is 30~mK.


\section{Photonic Links for Qubit Control}
A cryo-HFDD system (Fig.\ref{concept}), can be represented as a photonic control link (Fig.\ref{opticallink}) involving microwave signals carried over single mode fibers (SMF) to be detected by silicon photonic integrated photodiodes and fed to the qubits. This approach necessitates electro-optic modulators to perform the upconversion to the optical domain, which takes place at room temperature, and photodetection which occurs inside the dilution fridge. Our discussion focuses on studying the tradeoffs of a cryogenic photonic control link at different temperature stages to  define its limitations and describe its potential for scalability.

\subsection{Room Temperature Electro-Optic Transmission}
\begin{figure}[!tp]
    \centering
    \subfloat[]{\includegraphics[width=1.6in]{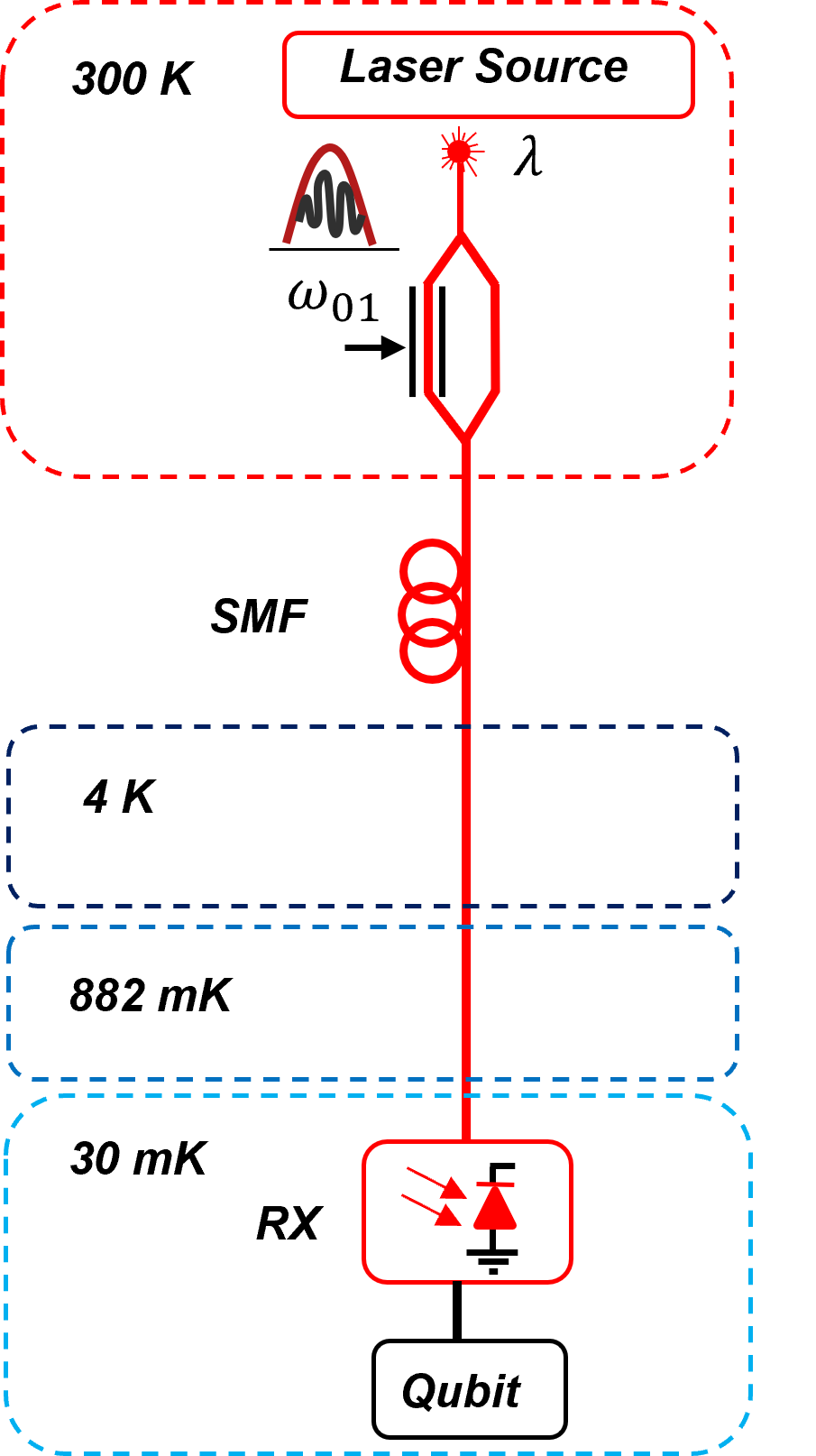}\label{<figure1>}}
     \subfloat[]{\includegraphics[width=1.57in]{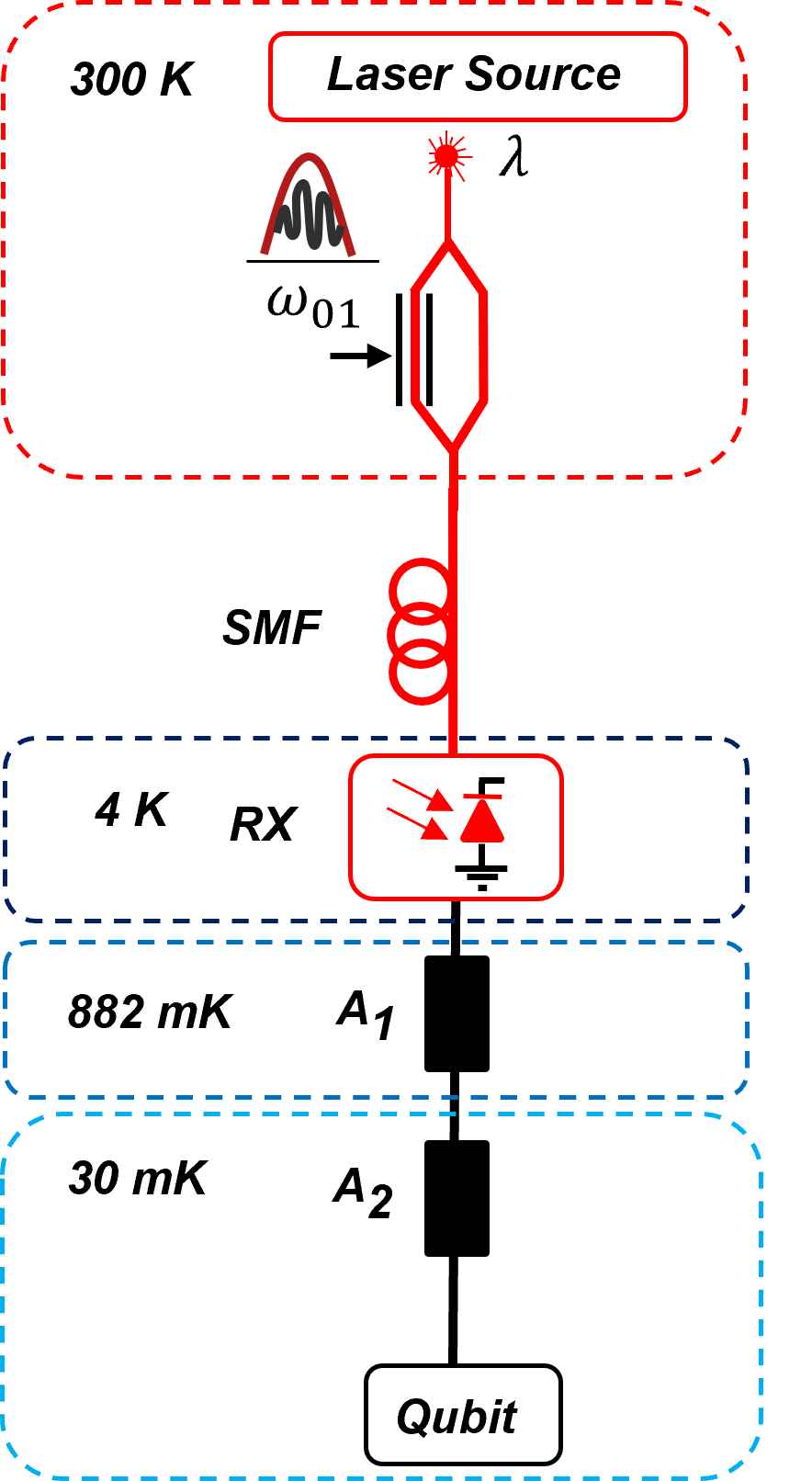}\label{<figure2>}}
     \caption{Photonic control architecture for a single qubit at (a) 30~mK, (b) 4~K.}
     \label{opticallink}
\end{figure}

Qubit microwave pulses (Gaussian, Raised Cosine, etc.), at a center frequency ($\omega_{01}$), are upconverted around the optical frequency ($\omega_{opt}$), typically over optical communication bands (C-band or O-band), with electro-optic modulators (EOMs), such as a dual-arm Mach-Zehnder Modulator (MZM), as shown in Fig.~\ref{opticallink}. The pulses are then downconverted in the fridge using photodetectors. Assuming a lossless MZM, we can represent the output optical power as~\cite{Chang_2002}:
\begin{equation}
P_{opt,out}={{P_{opt}}}[1+cos(\frac{V(t)}{V_{\pi}})]
\label{intensity}
\end{equation}

\noindent ${P_{opt}}$ is the average optical power, $V(t)= V_{DC}+ v(t)$ is the time varying voltage applied to the MZM where $V_{DC}$ is the DC bias and $v(t)=v_{m}sin(w_{01}t)$. Under quadrature biasing ($V_{DC}=\frac{V_{\pi}}{2}$), and assuming a linear modulator, the output optical power can be approximated as:
\begin{equation}
    P_{opt,out}={P_{opt}}(1-\epsilon_{m}sin(\omega_{01}t))
\end{equation}

\noindent where $\epsilon_{m}=\pi\frac{v_m}{V_\pi}$ is the modulation depth. To study the link $SNR$, we consider the transmitter noise sources: The relative intensity noise (RIN) of the laser source and the room temperature thermal noise from the EOM. The RIN noise current is defined as $S_{I}^{RIN}(\omega)= 10^{-\frac{RIN(\omega)}{10}}{I_{DC}}^2$, where $RIN(\omega) = -150$ dB/Hz, which is a typical value for modern laser sources \cite{RIN}, and $I_{DC}$ is the average photocurrent. The EOM room temperature thermal noise current, which is associated with the voltage swing $v(t)$ on the EOM electrodes, is defined as $S_{I}^{EOM}= S_{v}^{Vdr}(\omega) \times (\pi\frac{{I_{DC}}}{V_{\pi}})^2$ ~\cite{NIST}.
$S_{v}^{Vdr}(\omega)= 4k_{B}TZ_{dr}$, $V{\pi=2}$ V \cite{wang2018integrated} is the EOM half-wave voltage, $Z_{dr}=50~\Omega$ is the EOM electrode impedance, and $T$= 300~K is the room temperature, ignoring additional noise from the EOM driver circuitry. In the following analysis in this paper, we assume $\epsilon_m= 1$, where both the DC optical power, ${P_{opt}}$, and the sideband qubit optical power, defined as $P_{opt,\omega_{01}}=\epsilon_m{P_{opt}}$,  are equal. In Section \ref{sec_linearity}, we discuss the relationship between heat load, $SNR$, and the optical power at the desired signal when nonlinearity is taken into account.

 \subsection{Cryogenic Photonic Detection}
 \subsubsection{30 mK RX}
\begin{figure}[!tp]
    \includegraphics [width=0.4\textwidth]{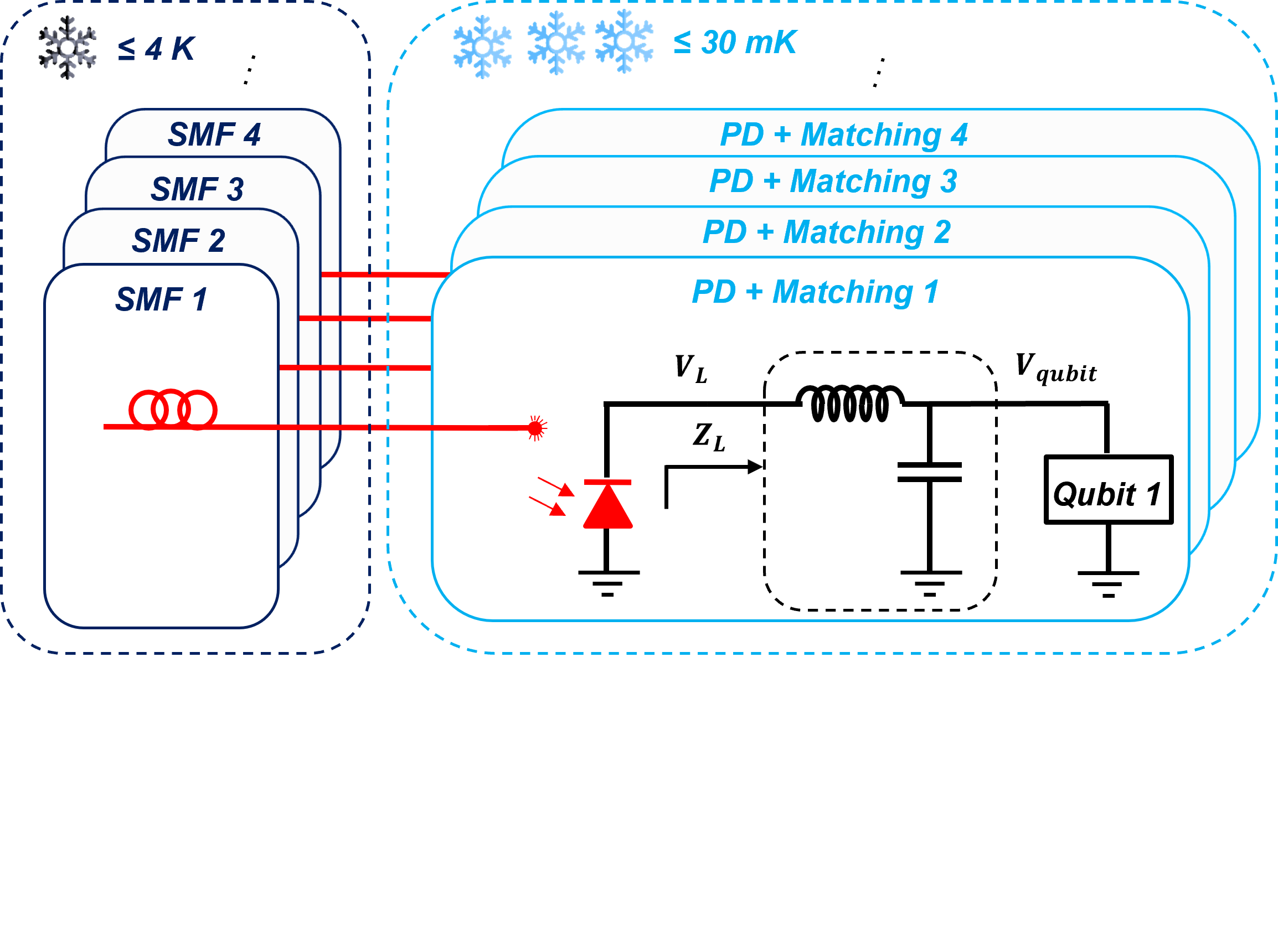}
     \centering
     \vspace{-45pt}
    \captionof{figure}{Cryogenic multi-qubit receiver at 30~mK with impedance matching.}
    \label{RXmk}
\end{figure}

We investigate the possibility of driving qubits immediately at 30 mK with a photonic link while meeting the required $SNR$ to achieve maximum gate fidelity. The receiver (Fig.~\ref{opticallink}~(a)) consists of photodetectors followed by matching networks (Fig.~\ref{RXmk}) to perform the optical-to-electrical downconversion. To individually control qubits with one fiber per photodiode, microwave signals at the qubit $\omega_{01}$ transition frequency are upconverted at optical wavelengths (C-band or O-Band) over a single-mode optical fiber (SMF) down to the MXC, as shown in Fig.~\ref{opticallink}~(a).  The fiber is then coupled to an on-chip photodiode with enough bandwidth to convert the signal carried by the optical wave to a microwave pulse at the qubit control frequency. 

The photodiode responsivity ($\mathcal{R}$) is an essential metric to determine heat load. Typical integrated photodetectors for most standard silicon photonic processes have Germanium (Ge) as the active area.  For an ideal photodetector with 100 \% quantum efficiency, the responsivity limit is $\mathcal{R}=\frac{q}{\hbar\omega_{opt}} \approx 1.25$ A/W at 1550 nm,
where $q$ is the electron charge, $\hbar$ is the reduced Planck constant, and $w_{opt}$ is the optical frequency. 
At cryogenic temperatures, the photodiode experiences a reduced responsivity due to the bandgap shift. Specifically, Ge photodiodes of a standard silicon photonic process were measured at O-band to have a responsivity $\mathcal{R}$ that ranges from 0.1 A/W to 0.15 A/W at 4~K, while at C-band, responsivity can drop below 0.1 A/W \cite{Responsivity}. Therefore, in the following discussion, we use a responsivity $\mathcal{R}= 0.1$ A/W for temperature stages $\leq~$ 4 K.


To fully calculate the link $SNR$, the receiver noise is incorporated. 
Here we only consider the shot noise of the photodiode, which is represented as $S_{I}^{Shot}(w)=2q{I_{DC}}$ \cite{Shotnoise}. Therefore, the optical link $SNR$ is defined as:   
\begin{equation}
\scriptsize
    SNR=\frac{V_{L}}{Z_{L}\sqrt{2q{I_{DC}} +10^{\frac{RIN(\omega)}{10}}{I_{DC}}^2+ 4KTZ_{dr}  (\pi\frac{{I_{DC}}}{V_{\pi}})^2}} 
    \label{Eq_SNR}
\end{equation}
where $Z_{L}=\frac{V_{L}}{{I_{\omega_{01}}}}$ is the load impedance presented at the photodetector (Fig.~\ref{RXmk}), $V_{L}=\frac{2 P_{\mu W}}{I_{\omega_{01}}}$ is the voltage at the photodetector load, {$I_{DC}$} = $ \mathcal{R} {P_{opt}}$ is the average photocurrent, $I_{\omega_{01}}=\mathcal{R} {P_{opt,\omega_{01}}}$ is the photocurrent at the qubit frequency, and $P_{\mu W}$ is the average microwave power needed to drive the qubit at the receiver stage. We here assume that for $\epsilon_m=1$, ${P_{opt}}=P_{opt,\omega_{01}}$, therefore $I_{DC}=I_{\omega_{01}}$.

The noise contribution of the receiver needs to be lower than the thermal noise floor at 30~mK to prevent its conversion into thermal noise from affecting the qubit. Assuming the $SNR$ is shot noise limited, $SNR$ $\propto$ $\sqrt{P_{DC}}$. To achieve the required $SNR$, the optical power must increase. 
This leads to a shot noise current increase, which necessitates transforming the 50~$\Omega$ qubit impedance ($Z_{qubit}$) to a much lower value at the photodiode output ($Z_{L}$). Therefore, a matching network (Fig. \ref{RXmk}) is needed to ensure the optimal $SNR$ while delivering the required microwave power to the qubit. 
In this analysis, we assume that the matching network is lossless, and the qubit input impedance does not contribute thermal noise. 
Therefore, the power delivered to the qubit is maintained, where the microwave power at the 30 mK stage can be defined as: 

\begin{equation}
    P_{\mu W,30mK}=\frac{V_{L}^2}{2Z_{L}}=\frac{V_{qubit}^2}{2Z_{qubit}}
\end{equation}

\begin{figure}[!htp]
    \includegraphics [width=3.4in]{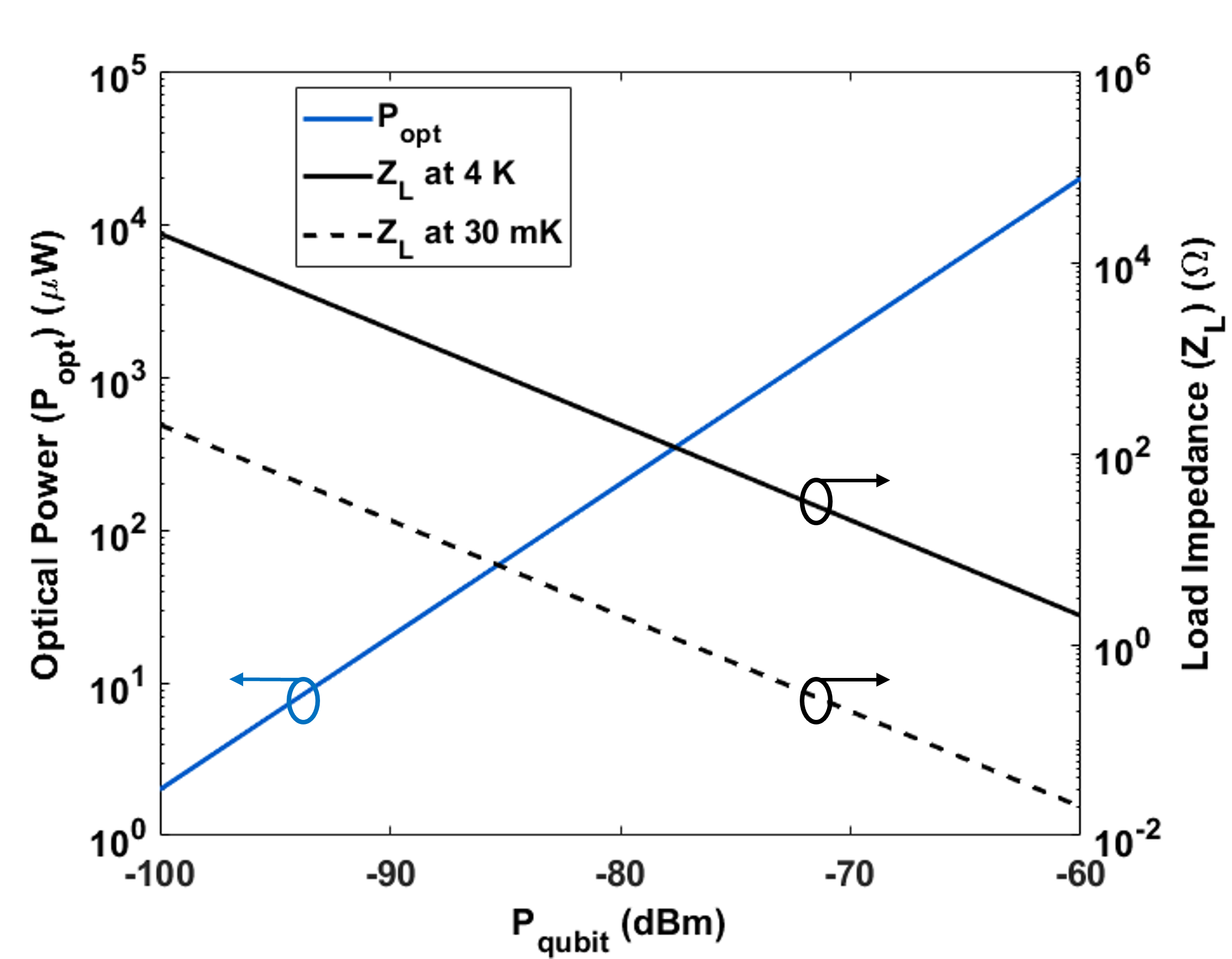}
     \centering
    \captionof{figure} {Minimum optical power $P_{opt}$ (blue) and load impedance $Z_L$ (black) to meet $SNR$ for different peak qubit power at 30~mK and 4~K.}
   \label{impedancepower}
\end{figure}

As shown in Fig.~\ref{impedancepower}, the minimum optical power (${P_{opt}}$) and photodiode load impedance ($Z_{L}$) can be optimized for each peak qubit power ($P_{qubit}$) to meet the $SNR$. For example, for $P_{qubit}$ = -70 dBm, the corresponding average qubit power ($P_{\mu W,30mK}$) is -80~dBm. 
It's important to note that in our analysis, -80~dBm average qubit power corresponds to -70~dBm peak power assuming a Gaussian pulse-shaped signal and 30\% activity (duty cycle)~\cite{Attenuation}. Therefore, $P_{opt,\omega_{01}}=100$~$\mu W$ is the signal optical power at the photodiode, and  $Z_{L}=$ 0.2  $\Omega$. This yields a heat load $P_{active}=200$ $\mu W$ per qubit  after accounting for a coupling insertion loss of approximately 3 dB \cite{Molnar}, assuming the photodiodes are built on a photonic integrated circuit (PIC) with edge coupling. 

The high optical power required renders this approach infeasible, given the limited cooling power of approximately 20 $\mu W$ at the 30 mK stage \cite{Attenuation}, along with the challenge of achieving practical impedance matching. Hence, we place the photodiodes at the 4~K stage for scalable qubit control to take advantage of the higher cooling power of around 1.5~W, as shown in Fig.~\ref{opticallink}~(b).

\subsubsection{4 K RX}
\begin{figure}[!tp]
    \includegraphics [width=3.4in]{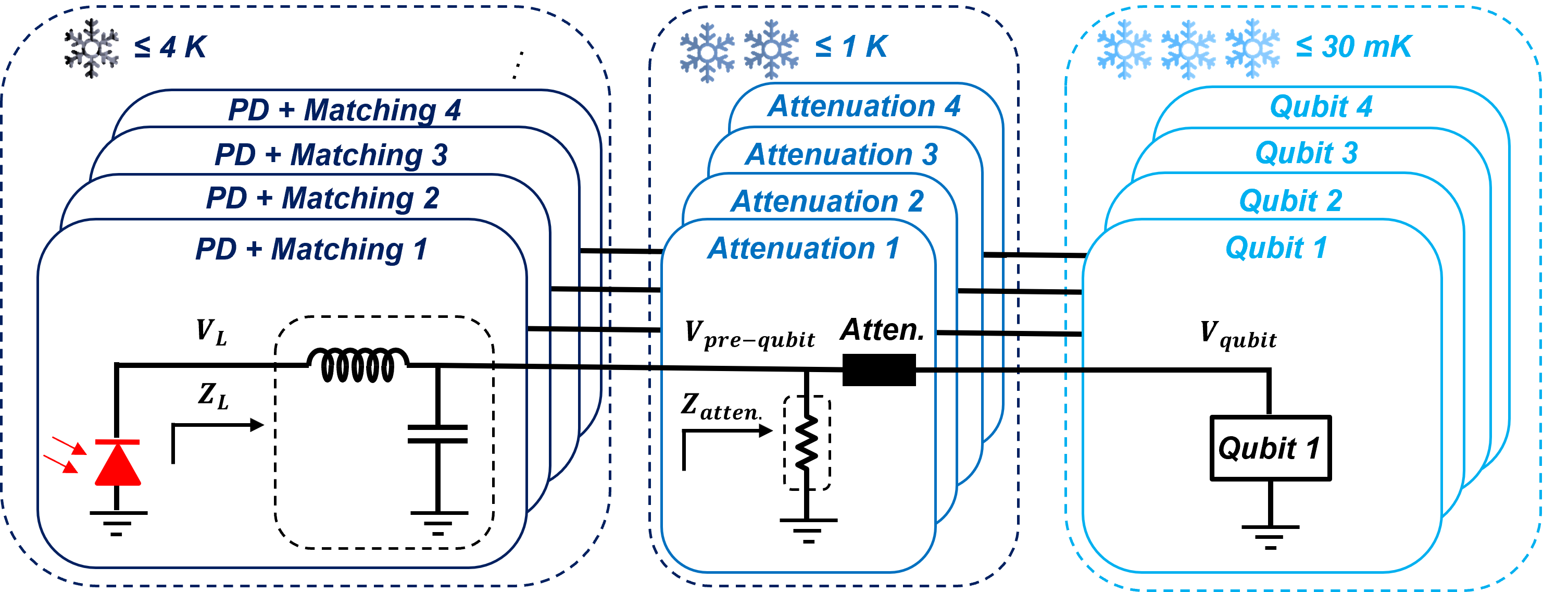}
     \centering
    \captionof{figure}{Cryogenic multi-qubit receiver at 4~K with impedance matching.}
    \label{RX4k}
\end{figure}

The fully passive multi-qubit optical receiver, shown in see Fig~\ref{RX4k}, is composed of an array of integrated photodetectors at 4 K. The signal is then filtered at the qubit frequency $\omega_{01}$ using a matching network, which acts as a microwave filter and optimizes the loading seen by the photodiode for best $SNR$. To define the receiver system specifications, we require that the noise generated by the receiver at 4~K be less than the noise floor of the 30~mK stage after 30 dB of attenuation ($A_1 + A_2$) between the two stages.  The link $SNR$ at the receiver output is calculated using Eq.~\ref{Eq_SNR}. Similar to the 30~mK RX case, the contributions of the RIN noise, EOM thermal noise, and the photodiode shot noise at 4~K are taken into account. Two key assumptions made here are that (i) the matching network after the photodiode is lossless, and (ii) the attenuators are placed at a lower temperature and the input impedance ($Z_{atten}$ = 50~$\Omega$) does not contribute any thermal noise at 4~K. We also assume 3 dB insertion due to the edge coupling to the PIC.

Fig.~\ref{impedancepower} shows the optical power ($P_{opt}$) needed to meet the minimum $SNR$. Since the same $SNR$ equation is applicable (Eq.~
\ref{Eq_SNR}), the needed optical power for the 4~K RX is the same as that for the 30~mK RX. However, the required impedance to match the photodiode output for best $SNR$ is 1000$\times$ that of the receiver at 30~mK due to the 30 dB attenuation between the two stages. 
At the end of this section, we analyze the effect of different attenuation on impedance matching. . 
For example, for the case of peak $P_{qubit}= $ -70 dBm ($P_{\mu W,30mK} = -80$~dBm), $P_{\mu W,4K} = -50$ dBm and $Z_L=200$ $\Omega$, where the average qubit power at 4~K is defined as:

\begin{equation}
    P_{\mu W,4K}=\frac{V_{L}^2}{2Z_{L}}=\frac{V_{pre-qubit}^2}{2Z_{atten}}
\end{equation}

\begin{figure}[!htp]
    \includegraphics [width=3.4in]{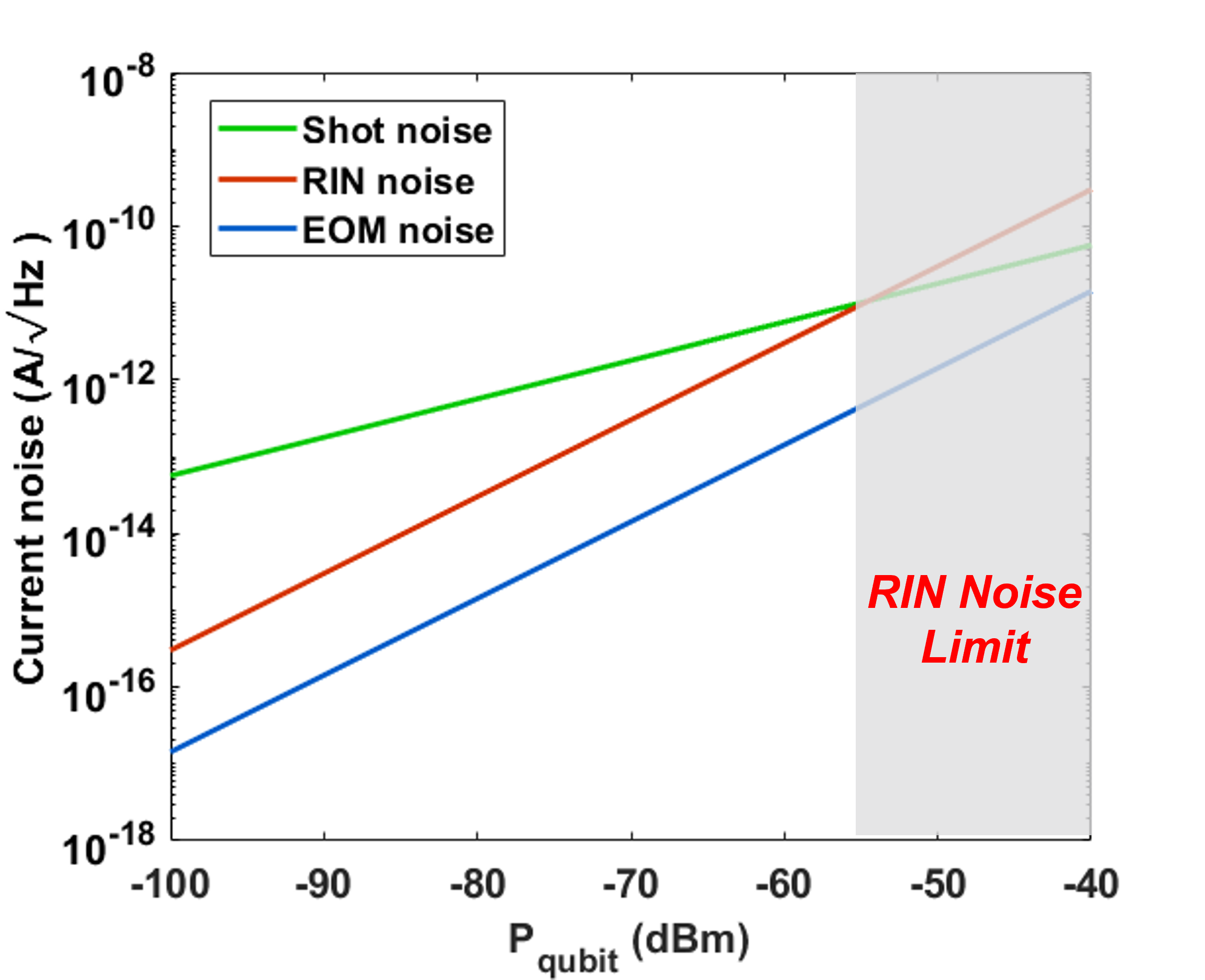}
    \centering
    \captionof{figure}{Current noise sources for different peak qubit power.}
    \label{currentnoise}
\end{figure}

\begin{figure}[!tp]
    \includegraphics [width=3.4in]{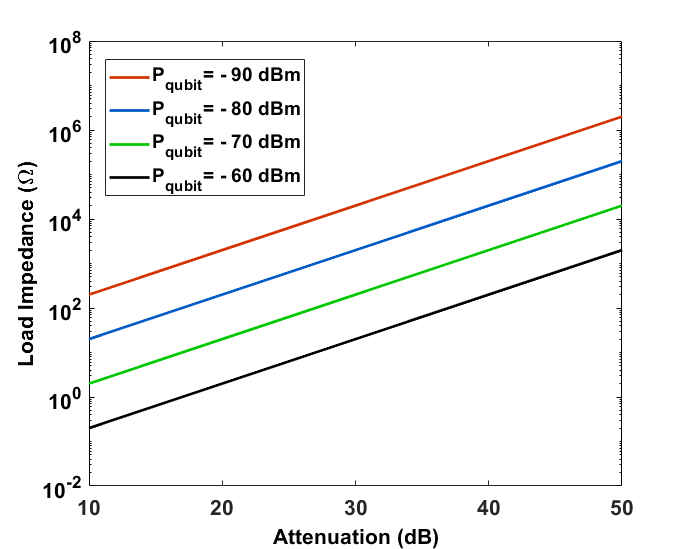}
     \centering
     \label{impedance_power}
    \captionof{figure} {Load impedance ($Z_L$) with respect to different attenuation levels for different peak qubit power.}
    \label{attimp}
\end{figure}

Fig.~\ref{currentnoise} shows the current noise sources with respect to the peak qubit powers ($P_{qubit}$). For $P_{qubit}\leq-55$ dBm, the link is shot noise limited. However, beyond this limit, the $SNR$ cannot be met since it is limited by the RIN noise. In addition, the required optical power ($P_{opt}$) increases to tens of mW. RIN noise suppression techniques \cite{nelson2008relative} can be implemented to reduce the required $P_{opt}$ above this limit. However, due to the complexity of the needed receiver, we do not consider these techniques. Furthermore, as mentioned above, the attenuation ($A_1 + A_2$) allows for controlling the required load impedance at the output of the photodiode ($Z_{L}$) as shown in Fig.~\ref{attimp}. 
This adds another degree of freedom to optimize the matching network and reduce its tunability requirements. 
For example, limiting $Z_L$ to a range that can be easily matched (20 $\Omega$ to 200 $\Omega$) is achievable by tuning the attenuation from 15 dB to 35 dB for peak $P_{qubit}$ ranging from -90 dBm to -60 dBm.

\subsection{Dense Channel Multiplexing \& Scalability}
\begin{figure}[!htp]
    \includegraphics [width=3.4in]{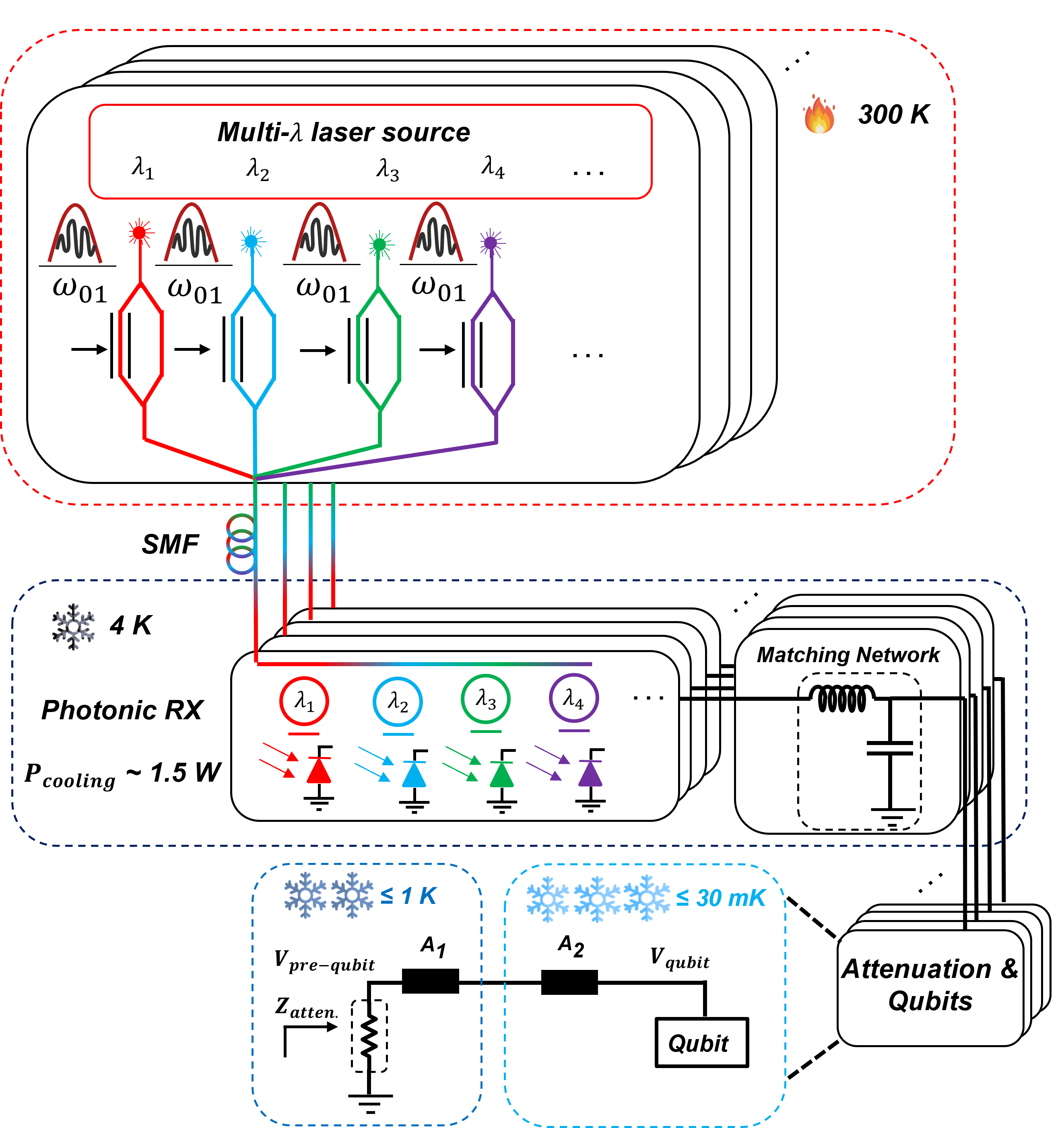}
     \centering
    \captionof{figure}{Dense cryogenic multi-qubit 4~K photonic WDM link.}
    \label{wdm}
\end{figure}

Wavelength division multiplexed (WDM) receivers, where multiple channels can be packed within a single fiber, can drastically increase link density \cite{Molnar}. In the context of qubit control, a WDM system enables a single fiber to act as the bus carrying multiple qubit data enveloped by frequencies across the optical communication band (Fig.~\ref{wdm}). The different channels are then separated via optical filters on the photonic RX. Each filter is followed by a photodiode to downconvert the pulse to the microwave domain, and a matching network to optimize the photodiode loading for maximum $SNR$. The signals are then sent to the attenuators and to the qubits at the lower temperature stages. The same $SNR$ analysis applies as the non-multiplexed receiver mentioned above. However, we estimate an additional 3 dB insertion loss due to the filtering \cite{Molnar}. For example, for a peak qubit power $P_{qubit}=-70$ dBm, $P_{active}=400$ $\mu W$, where $P_{opt,\omega_{01}}=100$ $\mu W$. 

From the above discussions about the $SNR$ requirements of the receiver, it is important for each optical filter to have high out-of-band rejection. For example, -80~dBm average RF power at the qubit necessitates an $SNR$ of 134~dB/Hz. This translates to 54~dB assuming $\sim$100~MHz bandwidth for the control of a fast qubit baseband Gaussian-shaped signal~\cite{sah2024decay}. Therefore, better than 60~dB of out-of-band rejection is needed to satisfy the required $SNR$ for most qubit powers. This will reduce interference from adjacent multiplexed channels. Optical filters such as cascaded ring resonators \cite{MRR} and Bragg gratings \cite{bragg} can achieve high extinction ratios and large free spectral ranges, which is highly desirable to maintain the minimum $SNR$ and maximize density.

\subsection{EOM nonlinearity impact on $SNR$ $\&$ heat load} \label{sec_linearity}
As mentioned above in Eq.~\ref{intensity}, the total output optical power of a MZM is composed of a component at the center frequency $\omega_{opt}$ of the laser source, and a sideband component centered at $\omega_{opt}\pm\omega_{01}$. The beating of both frequency components at the photodiode would result in a microwave pulse sent to the qubit. To assess the link performance ($P_{active}$ and $SNR$) with a nonlinear MZM, we need to estimate the maximum modulation depth $\epsilon_m$ at which the $SNR$ is achieved. By expanding Eq.~\ref{intensity}, the total output optical power of the MZM is expressed as:
\begin{equation}
 {{P_{opt}}}[1-sin(\pi\frac{v(t)}{V_{\pi}})] = P_{opt}[1-sin(\epsilon_msin(w_{01})t)]
                 \end{equation}
The output optical power at the sideband ($P_{opt,
\omega_{01}}$) changes with $\epsilon_m$.
The MZM transfer function is expanded as a series of Bessel functions of the first kind, with coefficients $J_n(\epsilon_{m})$, where $n$ represents the order of the Bessel components \cite{bridges2002distortion}. To formally characterize the nonlinear behavior, we consider two frequency tones $\omega_{1}$ and $\omega_{2}$ with the same modulation depth $\epsilon_m$, their relative sinusoidal modulation is expressed as:
\begin{equation}
    sin(\pi\frac{v_1(t)+v_2(t)}{V_{\pi}})=    sin(\epsilon_msin(\omega_{1}t)+\epsilon_msin(\omega_{2}t))
\end{equation}

\noindent Additionally, by applying the Jacobi-Anger expansion for Bessel functions, we can represent the previous equation as:
\begin{equation}
        \sum_{n=-\infty}^{\infty} \sum_{m=-\infty}^{\infty} J_n(\epsilon_m) J_m(\epsilon_m) \sin(n \omega_1 t + m \omega_2 t)
        \label{bessel_sum}
\end{equation}
where $n$ and $m$ are the Bessel function orders for both frequency tones \cite{cole2024equivalence}. We further expand this expression to a number of harmonics and intermodulation (IM) products that are squared to find the RF power after detection.

\begin{figure}[!htp]
    \includegraphics [width=3.4in]{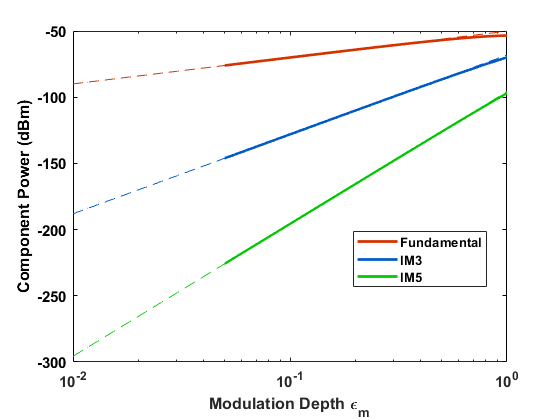}
     \centering
    \captionof{figure}{Microwave output power for the fundamental, IM3 and IM5 components at 4 K in terms of modulation depth for $P_{opt}=$ 100 $\mu$W calculated using Bessel functions expansion.}
    \label{Bessel}
\end{figure}

Since the modulator is assumed to be biased at quadrature, and referring to Eq. \ref{bessel_sum}, only odd harmonics are present. Fig. \ref{bessel_sum} shows the fundamental, IM3, and IM5 components at the output of the photodetector. The fundamental component is expressed as $\frac{1}{2}[(J_1(\epsilon_m)J_0(\epsilon_m)\mathcal{R}P_{opt}]^2Z_L$, where $P_{opt}=100$~$\mu W$, and $Z_L=200$ $\Omega$, which are optimum values for peak $P_{qubit}= -70$ dBm ($P_{\mu W,4K}=-50$ dBm).
The IM3 and IM5 components are proportional to $[(J_2(\epsilon_m)J_1(\epsilon_m)]^2$ and $[(J_3(\epsilon_m)J_2(\epsilon_m)]^2$ respectively.

 As mentioned previously, achieving $\geq$ 60 dB of dynamic range between the fundamental and IM terms (IM3 and IM5) is needed to satisfy the required $SNR$ for most average qubit powers. Considering third order nonlinearity for a regular MZM, this requirement can be met with $\epsilon_m < 0.15$, which yields $P_{opt}>25\times$ the previously calculated value of $100$~$\mu W$ for the average qubit power of -80~dBm to meet the same $SNR$. It should also be noted that for $\epsilon_m\neq1$,  $P_{opt,\omega_{01}} \neq {P_{opt}}$. The $SNR$ is then proportional to $\frac{I_{\omega_{01}}}{\sqrt{I_{DC}}}$, where  $I_{\omega_{01}}$ is approximately proportional to $\epsilon_mP_{opt}$ for small $\epsilon_m$. Therefore, $SNR$ $\propto$ $\epsilon_m\sqrt{P_{opt}}$, where $I_{DC}$ $\propto$ $P_{opt}$. This yields that the optical power ($P_{opt}$) scales with the square of the modulation depth ($\epsilon_m^2$). 

Existing linearization techniques such as ring-assisted MZMs \cite{RAMZI} and pre-distortion techniques \cite{okyere2017fifth} can linearize the third order nonlinearity. These modulators can extend $\epsilon_m$ to 0.5 to achieve the same $SNR$ which yields 4$\times$ the minimum $P_{opt}$. For total optical power to be reduced back to the minimum requirement, further efforts need to be made to achieve a linearized modulator with full modulation depth of $\epsilon_m = 1$. In this case, the optical power at the carrier will be the same as the power at the sidebands ($P_{opt}=P_{opt,\omega_{01}}$), minimizing the heat load in the fridge.

\section{Sub-THz Links for Qubit Control}
\begin{figure}[!htp]
    \includegraphics [width=2.9in]{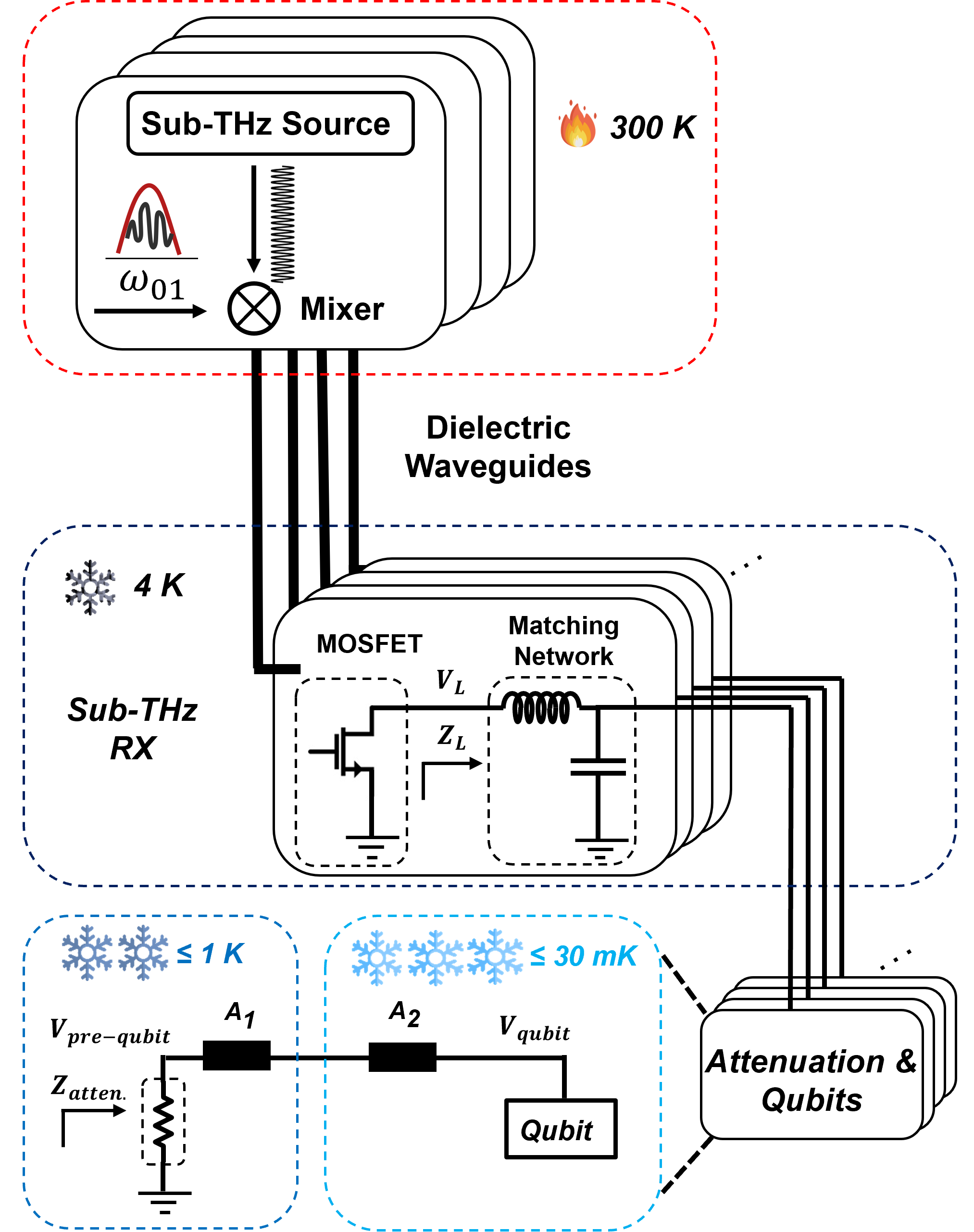}
    \centering
    \captionof{figure}{Multi-qubit sub-THz control link using cryogenic MOSFET detectors.}
    \label{thzlink}
\end{figure}

For a sub-THz link, the qubit pulses are upconverted over a carrier in the range of 140~GHz to 220 GHz, with room-temperature mixers, then fed to the dilution fridge with dielectric waveguides, as shown in Fig.~\ref{thzlink}. The waveguides are coupled to  cryogenic sub-THz CMOS chips using a chip-to-dielectric waveguide coupler~\cite{Jack, fukuda201112}. The signals then get downconverted with sub-THz MOSFET peak detectors \cite{ibrahim202029, khan2017nonlinear}. A matching network is then used to tune the load of the MOSFET detector for optimum $SNR$, and filter the signal to extract the qubit control pulses. These pulses are attenuated and sent to the qubits at lower temperature stages. Here, we present the analysis of a cryogenic sub-THz receiver at 4~K only, due to the fact that the required power at 30~mK is much higher than the available power budget (similar to photonic receivers at 30~mK).

The sub-THz transmitter and receiver both contribute to the total noise performance of the link. In the following analysis, we only consider the phase noise of the transmitter, and the shot noise at the receiver detector. In addition, we assume that the matching network is lossless and the attenuators do not contribute any thermal noise at 4~K. We also assume 30~dB attenuation between the 4~K and the 30~mK stages. The sub-THz MOSFET detector noise performance is dominated by the shot noise due to subthreshold and cryogenic operation \cite{khan2017nonlinear}. We assume a detector responsivity of $\mathcal{R}$ = 1 A/W at the 4~K stage~\cite{THzBackscatter}. 
Phase noise is estimated to be $PN(\omega)=-120$ dBc/Hz at 1 MHz offset from a 200 GHz source. This estimation is based on scaling the phase noise of a state-of-the-art Keysight PSG vector signal generator \cite{Keysight} with a frequency multiplier. The phase noise current is thus defined as $S_{I}^{PN}(\omega)= 10^{\frac{PN(\omega)}{10}}\mathcal{R}P_{sub-THz}= 10^{\frac{PN(\omega)}{10}}I_{DC}$. $P_{sub-THz}$ refers to the modulated sideband power around the carrier, and here we assume that the power at the carrier is equal to that at the sideband. The $SNR$ for the sub-THz link is therefore defined as:\\
\begin{equation}
\small         
    SNR=\frac{V_{L}}{Z_{L}\sqrt{2q{I_{DC}}  + (10^{\frac{PN(\omega)}{10}}{I_{DC}})^2}}
\end{equation}
where $Z_{L}=\frac{V_{L}}{{I_{\omega_{01}}}}$, $V_{L}=\frac{2P_{\mu W}}{{I_{\omega_{01}}}}$. For the same example of average qubit power of $P_{\mu W,30mK} = -80$ dBm, the needed sub-THz power $P_{sub-THz}$ is 10 $\mu W$ at the input of the MOSFET detector to meet the $SNR$ requirement of 134 dB, with $Z_L=200$ $\Omega$. It is important to note that it is difficult to achieve a frequency division multiplexing scheme due to complexity of implementing a bandpass filter in the sub-THz range, achieving  out-of-band rejection of $\geq$ 60~dB while supporting multiple channels. 

\begin{figure}[!htp]
    \includegraphics [width=3.4in]{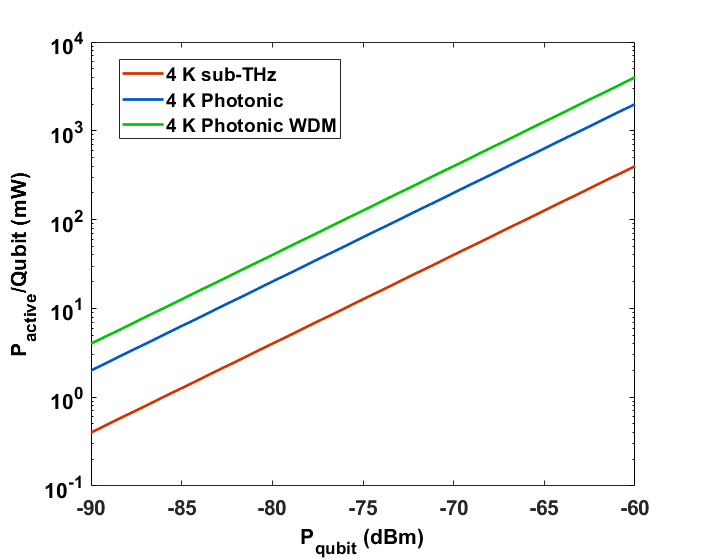}
    \centering
    \captionof{figure}{Heat load ($P_{active}$) of the control links at 4~K for different peak qubit power.}
    \label{linkcomp}
\end{figure}

Fig. \ref{linkcomp} shows the estimated heat load $P_{active}$ against different peak qubit powers. In this estimation, we account for 3 dB insertion loss for the chip-to-dielectric waveguide coupler \cite{Jack}. As discussed in Section \ref{section_passive_heat}, Teflon dielectric waveguides have an insertion loss of $\sim$5-10 dB/m at room temperature.
However, we estimate that the loss is $\sim$1-2 dB/m at 4~K, extrapolated from measurements at 18~GHz~\cite{jacob2002microwave}.
Here, we use 3 dB as a conservative estimate for the dielectric waveguide loss at 4~K. Hence, after accounting for the waveguide and coupler losses, for a peak $P_{qubit}=-70$ dBm ($P_{\mu W,30mK}=-80$~dBm), ${P_{active}}=40$ $\mu$W of power per qubit is consumed as heat at the 4~K receiver. Fig. \ref{linkcomp} also shows a comparison with the other photonic link modalities.

\begin{figure}[!htp]
    \includegraphics [width=3.4in]{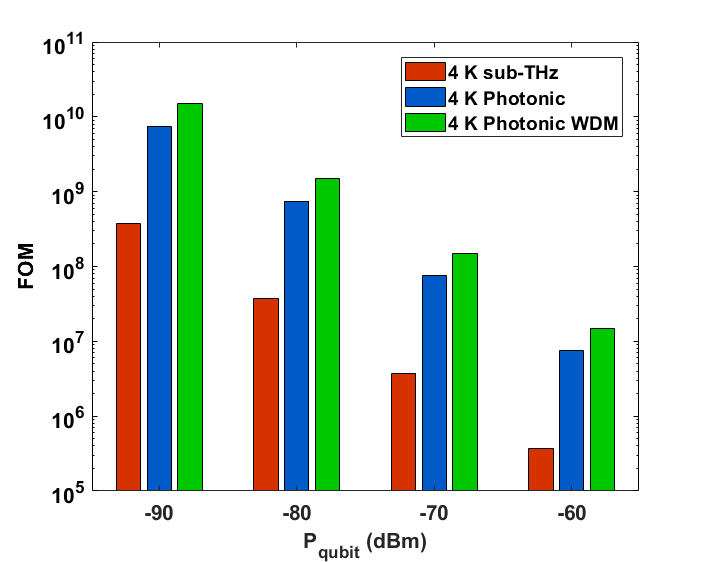}
    \centering
    \captionof{figure}{Figure of Merit (Eq.~\ref{FOM}) for different peak qubit power, where $N_{qubit}$ = 4 for the 4 K photonic WDM case and $N_{qubit}$ = 1 for the other two cases.}
    \label{density}
\end{figure}

While this control link benefits from low power consumption, the size of dielectric waveguides limits its scalability compared to optical fibers.
In the range of 140~GHz to 220~GHz, a 1~mm core size is needed compared to only 10~$\mu m$ for an optical fiber.
To better quantify the comparison with the photonic link, and take into account both power density and cable density, we define a figure of merit (FOM) such that:
\begin{equation}
\small         
    FOM=\frac{P_{cooling, ~4K}} {{P_{active}{~\times~\Delta L/N_{qubit}}}}
    \label{FOM}
\end{equation}
where $P_{cooling,~4K}$ is the cooling power at 4 K, estimated to be 1.5 W,
$P_{active}$ is the active heat load at the stage per qubit, and $\Delta L/N_{qubit}$ is the pitch size between cables with respect to the number of qubits driven in that cable. As shown in Fig.~\ref{density}, a wavelength division multiplexed photonic link at 4~K outperforms other control links based on our defined FOM. In this comparison, we assume $N_{qubit}=4$ for the WDM photonic link.

\section{Future scalability projections}
\begin{figure}[!htp]
    \includegraphics [width=3.4in]{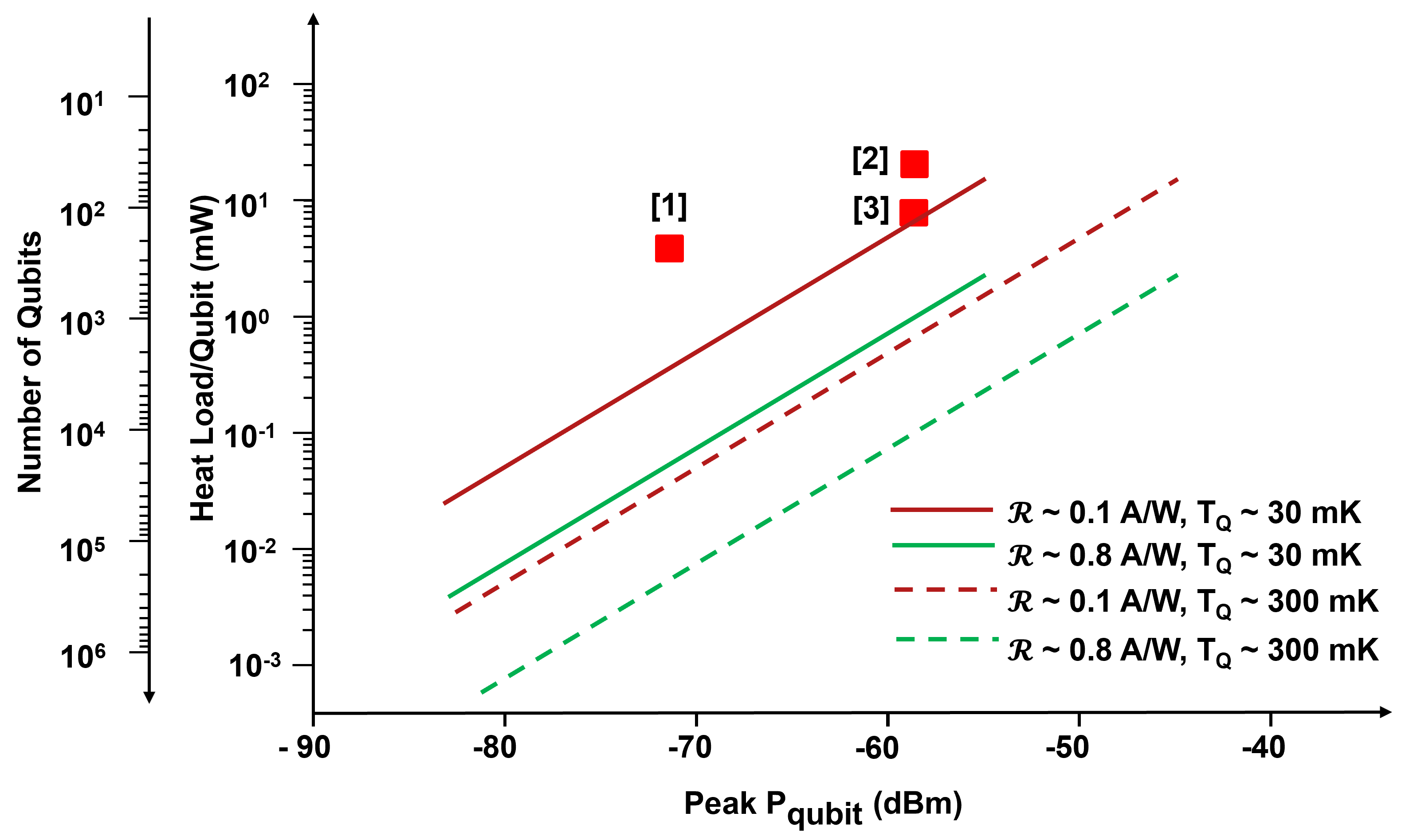}
    \centering
    \captionof{figure} {Cryogenic 4 K photonic all-passive WDM link future outlook in comparison with current state-of-the-art cryoCMOS controllers for superconducting qubits. $T_Q$ is the qubit operating temperature and $\mathcal{R}$ is the photodiode responsivity.}
    \label{outlook}
\end{figure}

Fig. \ref{outlook} shows the heat load ($P_{active}$) per qubit and the projected number of qubits for the proposed all-passive WDM cryo-HFDD architecture in comparison with the state-of-the-art cryoCMOS pulse modulation controllers. Based on our previous analysis, and with a photodiode responsivity $\mathcal{R}=0.1$~A/W, photonic links with wavelength division multiplexing can aggressively scale quantum computing systems in the near-future to $\sim$ 7500 qubits for $P_{qubit}$ = -70~dBm \cite{Google,Attenuation}, given the current state-of-the-art dilution fridges cooling budget of 1.5~W at 4~K. 
This can be achieved using 1875 fibers with a multiplexing factor of 4. 
Achieving the $SNR$ specification we define here yields more than 99.999 \% gate fidelity.  The gate error of a 20 ns long $\pi$ pulse ($\Delta t_{gate}$)\cite{Google} for a shot-noise limited control system is defined as $P_{error}= (\frac{\pi}{2})^2\frac{1}{N_{tot,opt}}\propto~ \frac{1}{(SNR)^2}$ where $N_{tot,opt}=\frac{\Delta t_{gate}P_{opt}}{\hbar\omega_{opt}}$ is the total number of optical photons in the gate pulse and $\Delta t_{gate}=20$ ns is the gate duration \cite{Amir}.
Furthermore, improving the photodiode responsivity enables the control of a larger number of qubits. For example, by extending the photodiode bandgap \cite{NIST_2} to reach room-temperature responsivity values of 0.8 A/W, it is possible to reach near-term scaling to $\sim$ 30000 qubits. Additionally, by increasing the qubit operating temperature ($T_Q$) to $\sim$ 300 mK with mm-wave qubits~\cite{anferov2025millimeter, Amir}, the $SNR$ requirement is further relaxed due to the increased thermal noise floor at 300~mK compared to 30~mK. The proposed control link can therefore enable the control of hundreds of thousands of qubits thanks to the passive matching network approach that could optimize the loading of the detectors to maximize $SNR$. In the future, more advanced refrigeration systems could achieve 1~KW of cooling power at 4~K~\cite{1KW_Cooling_Linde}, enabling the control of millions of qubits.

\section{Conclusion}
Conventional coaxial cables for controlling qubits have proven to limit scalability to hundreds of qubits. 
Cryo-HFDD architecture on the other hand has the potential to break through this barrier. Specifically, we analyze the link design tradeoffs of photonic and sub-THz control architectures such as receiver and qubit operating temperature, system attenuation, component losses, detector responsivity ($\mathcal{R}$) and loading, and cable density. Multiplexed photonic links stand out as a largely superior communication modality to overcome the current scalability limitations of coaxial cables. With further improvements to electro-optic modulator linearity, photodiode responsivity, qubit operating temperature and cooling budget at 4~K, the proposed cryo-HFDD platform opens the door to highly scalable quantum information processing.

\bibliography{references}
\end{document}